# Identify the Beehive Sound using Deep Learning

Shah Jafor Sadeek Quaderi[1], Sadia Afrin Labonno[2],
Sadia Mostafa[3] and Shamim Akhter[4]

[1]Department of Computer Science, Faculty of Computer Science and Information
Technology, University of Malaya, Kuala Lumpur, Malaysia
[2, 3]AISIP Lab, Department of Computer Science and Engineering, International
University of Business Agriculture and Technology, Dhaka, Bangladesh
[4]Department of Computer Science and Engineering,
Stamford University Bangladesh, Dhaka, Bangladesh

## Abstract

*Flowers play an essential role in removing the duller from the environment. The life cycle of the flowering plants involves pollination, fertilization, flowering, seed- formation, dispersion, and germination. Honeybees pollinate approximately 75% of all flowering plants. Environmental pollution, climate change, natural landscape demolition, and so on, threaten the natural habitats, thus continuously reducing the number of honeybees. As a result, several researchers are attempting to resolve this issue. Applying acoustic classification to recordings of beehive sounds may be a way of detecting changes within them. In this research, we use deep learning techniques, namely Sequential Neural Network, Convolutional Neural Network, and Recurrent Neural Network, on the recorded sounds to classify bee sounds from the non- beehive noises. In addition, we perform a comparative study among some popular non-deep learning techniques, namely Support Vector Machine, Decision Tree, Random Forest, and Naïve Bayes, with the deep learning techniques. The techniques are also verified on the combined recorded sounds (25-75% noises).*

## Keywords

## 1. Introduction

Ensuring the long-term sustainability of natural ecosystems, monitoring honeybee colonies and identifying their locations is considered significant research. Hence, many researchers prefer to conduct their research in that field. The audio-based technique is denoted as an efficient technique for identifying honeybee colonies as well as beehive sounds [1-3]. The initial step toward developing audio-based beehive monitoring technology is to develop systems that can distinguish between bee and non-bee sounds may be collected. Non-bee sounds are often related to nearby surroundings and occurrences around the hive, including urban, animals, rain, and maintenance sounds. Thus, this research automatically detects given audio recordings obtained inside beehives. In comparison to beehive sounds, the group of non-beehives noises may be shorter in time duration, which effectively differentiates those sounds.

In this paper, we initially preprocess all the beehives sound waves and separate the bee and non-bee waves with two seconds duration. After that, the waves are processed to extract 134 features, including Chromagram STFT (provides details about the pitch category and signal pattern), Root-Mean-Square (RMSE) Energy, Spectral Centroid, Spectral Bandwidth, Spectral Rolloff, Zero Crossing Rate, and Mel Frequency Cepstral Coefficients (MFCCs) which is comprised of 128 individual sequences (mfcc1 to mfcc128). We apply feature selection methods, namely Correlation Coefficient and Select_K_Best, on 134 features and identify the most useful 26 features. These 26 feature-based datasets are used to train the deep and non-deep learning classification methods. The testing accuracy of the deep learning techniques are varied from 85.04-99.26%, and non-deep learning techniques are varied from 88.06- 97.74%. In addition, five different sounds (beehive sounds and 25-75% additional non-beehive sounds) are extracted from source wave file randomly to justify this research.

This section briefly provides information and introduces the proposed research on beehive sound identification. Section- 2 presents the related works on bee sound detection and classification. Section- 3 discusses the methodology of the proposed research with a detailed explanation of datasets, features extraction procedures, and their implementations on different ML techniques. Section- 4 discusses the results of classification techniques and validation methods with combined recorded datasets. Section- 5 explains limitations and future work. Finally, we end up this work in section 6 with a conclusion.

## 2. RELATED WORK

The sound of a beehive is a mixture of the individual contributions of sounds generated from each bee colony; thus, we prefer to examine the recording of our dataset beehive sounds into the frequency domain. To identify beehive sounds, researchers used the Mel-frequency Cepstral Coefficients (MFCCs) as important features [4]. Also, the denoising techniques and useful features, namely Wavelet transform and characteristics, chroma, and spectral contrast were used to determine beehive sounds. Besides, the beehive sound analysis researches comprise data preprocessing, hand-crafted features, and domain knowledge in order to clean the recordings from non-beehive sounds. For filtering the acoustic signal, the authors of [5] used a Butterworth filter with cut-off frequencies of 100 Hz and 2000 Hz and excluded non-bee sounds that were not predicted to belong to the bee sound class. The research [6] clearly distinguishes three types of sounds- beehive, ambient, and cricket. Machine learning approaches, particularly deep learning methods, can reduce the number of handmade characteristics and domain knowledge introduced by bias and limit the modeling capabilities of sound identification algorithms to a degree. The paper [7] demonstrated that DNNs outperformed shallower methods like Gaussian mixture models in environmental sound scene analysis (GMMs) where normal acoustic classification technique [8] or IoT-based bee swarm activity acoustic classification technique [9] could a notable way to record that sound through hand-crafted attributes [10]. In [11], the authors created an annotated dataset from beehive sound recording.

Two machine learning approaches, namely, SVM [12, 13] and CNN [14, 15], were considered as beehive sound recognition systems. Besides, in [16], there is a wide comparison of the impact of CNN and machine learning classification algorithms on beehive sound recognition. Nevertheless, the machine learning classification techniques are preferred for monitoring the beehive [17] as well as identifying beehive sounds through the data collected via IoT sensors [18]. The machine learning classification techniques- Random Forest (RF) and Support Vector Machines (SVM) are regarded as the most employed classifiers non-deep learning approaches for bees buzzing-sounds classification [19], while the MFCC is considered the most used feature extraction strategy [20, 21]. Further- more, Decision Tree (J48) and Naïve Bayes are also effective techniques for identifying to predict the bees' sound based on their flying sound [22].

So far on the above-mentioned researches the feature engineering, including feature extraction procedure and selecting the best features for classification which is missing. Since many features could be extracted from the beehive audio; therefore, it seems too essential to find out the convenient features for beehive sound identification. Also, to make compatible with chosen techniques for obtaining significant outcomes, it is obvious to identify the useful features through feature Engineering. The implementation and testing of machine learning techniques were also done without the ideal combination of parameters. Thus, parameter tuning or application grid search for optimum parameter combination is also missing. The test datasets only tested classification accuracy. In our proposed implementation, we consider feature engineering approaches on the datasets, optimum combination of parameters in model architecture, and model validation or workability by using combined sounds as well.

## 3. METHODOLOGY

This research followed a customized framework for developing a model to classify bee sounds from the non-beehive noises through deep learning and non-deep learning approaches (machine learning algorithms) and from the online datasets. The framework comprises four stages (Figure 1); the initial stage describes the online data collecting information. The next stage highlights the datasets' preprocessing by a couple of procedures. Following the preprocessing, the feature extraction- extract the available features from the dataset, and feature selection- identify the most suitable features. The stage is for implementing preferred machine learning techniques. Figure 1 demonstrates the methodology in this research.

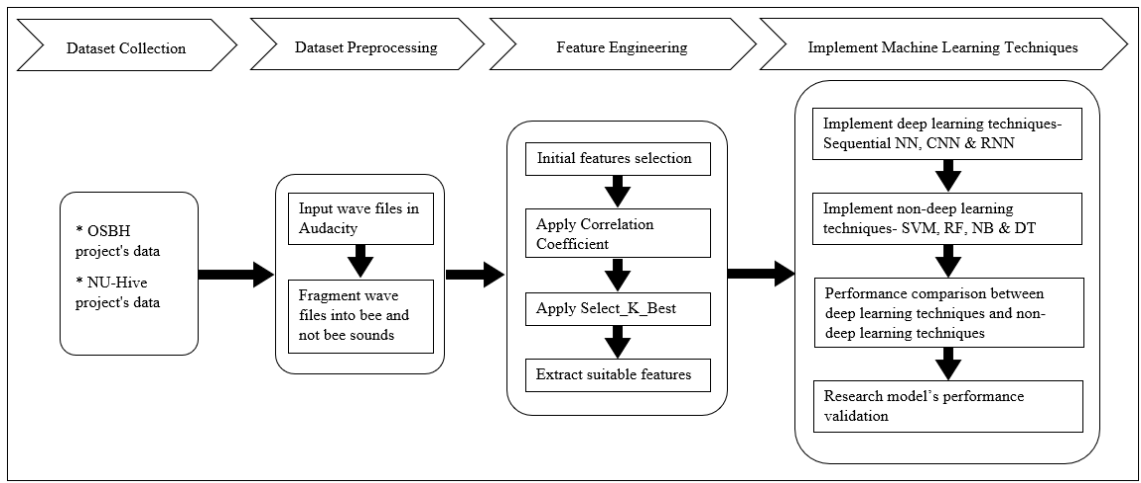

Figure 1. Methodology of this research

### 3.1. Dataset Collection

For developing this research framework, it is obvious to collect the dataset from a reliable source and research. As a result, we obtain datasets from two projects, namely the Open-Source Beehive (OSBH) and the NU-Hive [11]. The primary objective of those projects is to develop beehive surveillance systems capable of identifying and predicting hive's state.

### 3.2. Dataset Preprocessing

In the earlier research [11], Sonic Visualiser is utilized to finish the preceding research's sound annotation tasks. Through labeling the whole recording with these pairs of instances matching to the start and end of external sound periods, the entire recording was divided into Bee and noBee intervals and save those as text files. The noBee intervals denote times during which an external sound is audible. An example of this process is shown in Figure 2.

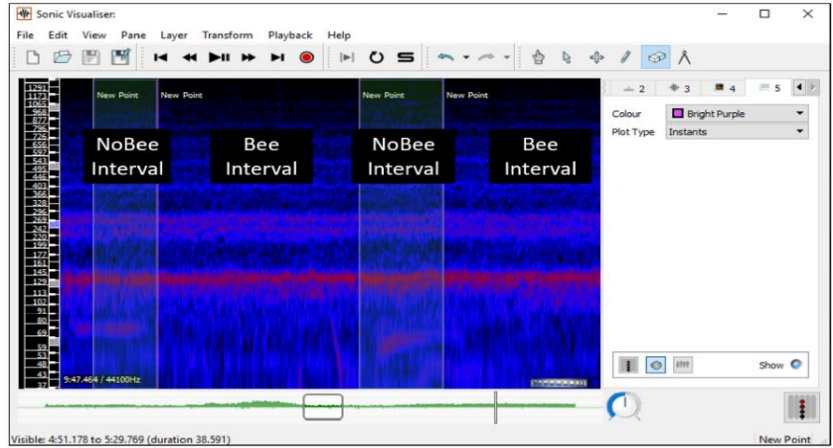

Figure 2.  Annotation procedure in the earlier research for one audio file by Sonic Visualiser3

However, for this research, we prefer Audacity software to traverse in the wave files (in the dataset); those wave files are categorized on bee and non-bee sounds in the dataset. Based on the given annotation, fragment them into two-second long wave files (either bee or not bee). We extract 4070 wave files in total where the number of bee sound and non-bee sound clips are 1100 and 2970, respectively. An example of a waveform in this process is shown in Figure 3.

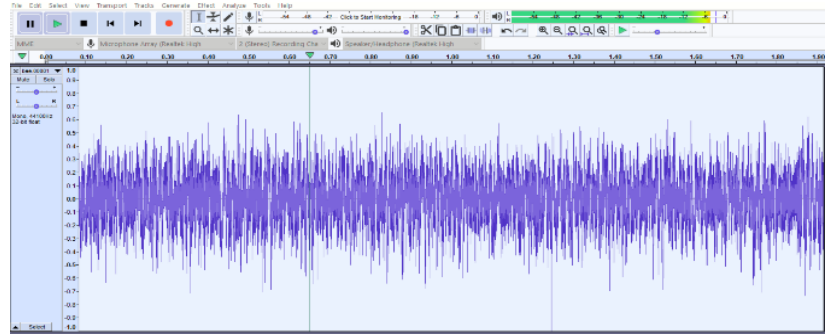

Figure 3.  Annotation procedure in the current research for one audio file by Audacity

The audio files are processed at a sampling rate of 22050 Hz and divided into two-second blocks. Segments whose lengths are less than the specified block length is completed by repeating the audio signal until the required block length is fulfilled. A label is given to each block depending on the current annotations. If the whole segment does not include or any external combined sound interval, the label Bee is assigned. Similarly, if at least a portion of the segment includes an external sound event, the label noBee is applied. Figure 4 highlights a sample amplitude from the combined form of Bee and noBee sound.

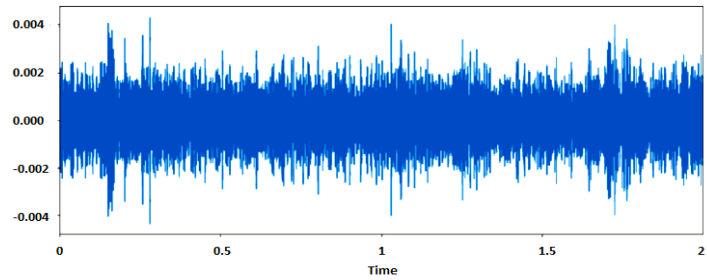

Figure 4. Sample amplitude from the combined form of Bee and noBee sound

Sound frequencies are shown in a spectrogram, which is a visual representation of how they change over time. It is a visual representation of how frequencies change over time for a given wave file. Figure 5 shows the spectrogram of that combined amplitude.

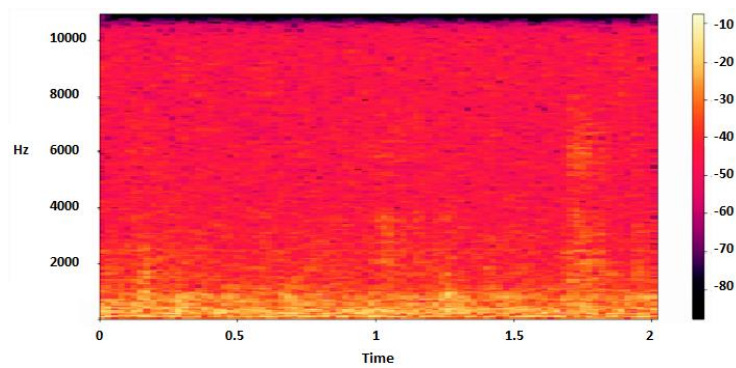

Figure 5. Spectrogram of the sample amplitude

Each segment wave file (Bee and No Bee) is converted to spectrogram separately. Using STFT, we are able to determine the amplitude of a particular frequency at a given time by converting data into signal. It is possible to determine the amplitude of different frequencies in an audio stream by using this Short-time Fourier transform. Different Mel frequency Cepstral Coefficients may be measured from the audio source using this Fourier transformation. Figure 6 demonstrate the MFCCs of that sample amplitude.

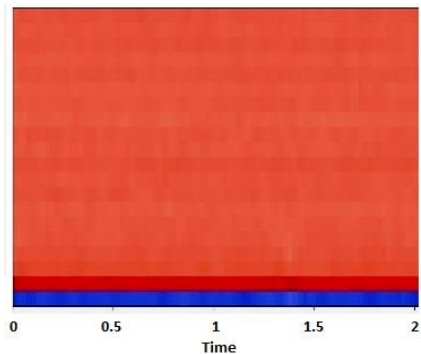

Figure 6. MFCCs of the sample amplitude

## 3.3. Feature Engineering

Features are some valuable characteristics to describe a large set of data accurately. At first, we need to use extract the suitable features from the data set and apply some procedure to separate the most consistent, non-redundant, and relevant features to use in machine learning technique construction. Thus, feature engineering involves two naming steps-namely feature extraction and suitable feature selection.

### 3.3.1.  Initial Features Extraction

A large number of features - 134 features - are extracted at the initial phase of this feature engineering process. This wide feature set comprises Spectral Centroid, Spectral Bandwidth, Chromagram, Zero Crossing Rate, Roll Off, Root Mean Square Error and Mel Frequency Cepstral Coefficients (of 128 different sequences).

The Spectral Centroid of a sound represents the "center of mass" of that sound, and it is derived by taking the sound frequencies' average weight (in Equation. 1). If the frequencies of a sound are consistent continuously, the Spectral Centroid would've been located around its center; conversely, if the sound has high frequencies at its ending portion, the centroid might be found near its end. The centroid measures spectral shape, and relatively high centroid values relate to "brighter" timbral textures with a greater proportion of high frequencies. For the purposes of this equation, M[n] denotes the value of the Fourier transform at time frame t and frequency bin n.

$$C_t = \sum_{n=1}^{N} n M_t[n] / \sum_{n=1}^{N} M_t[n] \quad (1)$$

The Spectral Bandwidth is calculated from the Spectral Centroid. It provides information about the audio signal's variance relative to the Spectral Centroid. The Spectral Bandwidth has a strong connection with perceived timbre in terms of perceived timbre. The bandwidth is proportionate to the amount of energy that is spread across different frequency bands. Mathematically, it is the mean weight of the distances between frequency bands and the Spectral Centroid.

Chromagram is precisely correlated audio signal's pitch. A pitch may be divided into two categories, which are referred to as tone chroma and height. The tone height relates to the octave number; besides, the chroma is associated with pitch spelling characteristics. For depicting the pitch data of the signals, we con- ducted Chromagram STFT to acquire the chroma characteristics. The primary concept of chroma features is to collect all spectral information associated with a specific pitch class into a single coefficient. According to its audio octave, the audio stream might be categorized into 12 different pitches. Chromagram projects the whole spectrum into 12 pitches.

The Zero Crossing Rate notifies a signal that is changing from the polar position of the horizontal axis. The information provided by time-domain zero-crossings is utilized to quantify the noisiness of the signal, and the mean and variance of zero crossings over the timeframe in the texture window are used as features. A signal's ability to transition from positive to zero to negative and negative to zero to positive in a certain period is measured by this parameter. Equation 2 depicts Zero Crossing Rate's equation throughout a time-domain that has been specified.

$$Z_t = 1/2 \sum_{n=1}^{N} |sign(x[n] - sign(x[n-1])| \quad (2)$$

Roll Off indicates the limits of high treble and the low bass in a frequency response curve. Furthermore, the frequency Rt denotes the Roll Off, and it concentrated 85% of the magnitude distribution-$\sum_{n=1}^{R_t} M_t[n] = 0.85 \sum_{n=1}^{R_t} M_t[n]$. Multiple features are created by taking the variance and mean of the Roll Off over the time-frames within the texture window.

Furthermore, RMSE another notable features for audio signal classification research. It represents the Root-Mean-Square energy value of the wave signal as a feature. Root Mean Square Error uses to characterize the average of continuous varying audio signals. Its value is calculated as the frame-wise from the audio signal.

Mel Frequency Cepstral Coefficients are popular features in the sound recognition process. MFCCs include the data of different spectrum bands' rate changes. According to the Fourier transform, MFCCs are considered as perceptually determined features. After obtaining the Fourier transform of an analysis window, the magnitude spectrum is processed through a Mel filterbank with dynamic bandwidth emulating the human ear, i.e., short bandwidth at a lower frequency and broad bandwidth at a higher frequency. The generated energy from the filterbank is log-transformed, and MFCCs are formed by the Discrete Cosine Transform of the generated outputs. In addition, the Python library of MFCC function can generate maximum 1 to 128 sequences of MFCC based on requirement.

### 3.3.2. Suitable Features Selection

Throughout our analysis, we realized that it is a time-consuming process to execute machine learning techniques on 134 features. Thus, two feature selection mechanisms – Correlation Coefficient and Select_K_Best – are used to extract the most suitable features. Kendall's Tau formula has been used to conduct the Correlation Coefficient test among the features (in Equation 3). Table 1 shows the Correlation Coefficient of the selected features. Besides, the Select_K_Best has computed by Python's f_classif function where this function ranks k number of the feature by ANOVA's F value; the formula of F Value ANOVA calculation is shown in Equation 4.

$$Kendall's\ Tau\ Correlation\ Coefficient = \frac{C-D}{C+D} \quad (3)$$

C is the number of concordant pairs and D is the number of discordant pairs.

$$F\ Value = \frac{SSE1 - SSE2/m}{SSE2/n - k}(4)$$

SSE = residual sum of squares, m = number of restrictions, k = number of independent variables and n= umber of observations.

Based on the correlation coefficient analysis, the correlation coefficient scores with the target variable are decreased (more negative values) after the mfcc20. Hence, we discard rest MFCCs after mfcc20. Table 1 represents a shortlist of the correlation coefficient score with the target variable (label) in descending order (with mfcc75 and mfcc128 correlation coefficient scores additionally).

Another feature selection mechanism- Select_K_Best is applied to determine the most convenient features for this research. Select_K_Best's f_classif function receives a score function as an argument that must be applicable to a pair of features. That scores should be provided as an array, one for each feature; then it picks the first k features from the selected features with the highest scores and generates a ranking of all features based on obtained scores. Table 2 shows the Select_K_Best scores of the selected features with the target variable.

Table 1. Correlation coefficient value with target variable.

| Features | Correlation Coefficient Value with Label | Features | Correlation Coefficient Value with Label |
|---|---|---|---|
| spectral_bandwidth | 0.492264 | mfcc5 | -0.152646 |
| rolloff | 0.490877 | mfcc12 | -0.296329 |
| spectral_centroid | 0.488098 | mfcc13 | -0.394607 |
| zero_crossing_rate | 0.456123 | rmse | -0.423690 |
| mfcc19 | 0.430004 | mfcc1 | -0.435055 |
| mfcc18 | 0.368745 | mfcc2 | -0.474572 |
| mfcc3 | 0.366760 | mfcc4 | -0.478908 |
| mfcc20 | 0.358174 | mfcc7 | -0.484232 |
| mfcc17 | 0.307512 | mfcc8 | -0.490135 |
| chroma_stft | 0.064920 | mfcc14 | -0.495657 |
| mfcc11 | 0.061819 | mfcc6 | -0.499427 |
| mfcc10 | 0.052450 | mfcc21 | -0.538421 |
| mfcc16 | -0.011519 | mfcc22 | -0.591179 |
| mfcc9 | -0.077382 | mfcc75 | -0.571325 |
| mfcc15 | -0.110117 | mfcc128 | -0.697854 |

Table 2. Select_K_Best scores with target variable.

| Features | Select_K_Best Scores with Label | Features | Select_K_Best Score with Label |
|---|---|---|---|
| rolloff | 2446.838434 | mfcc20 | 655.009811 |
| spectral_bandwidth | 2218.814429 | mfcc12 | 526.089194 |
| mfcc14 | 2016.781313 | rmse | 522.954252 |
| spectral_centroid | 1956.170643 | mfcc17 | 480.911989 |
| mfcc6 | 1674.402224 | mfcc5 | 458.215400 |
| mfcc8 | 1560.042411 | mfcc15 | 426.132326 |
| mfcc2 | 1538.185904 | mfcc9 | 410.707064 |
| mfcc7 | 1457.515419 | chroma_stft | 398.841436 |
| mfcc4 | 1454.056414 | mfcc16 | 378.422331 |
| zero_crossing_rate | 1288.146314 | mfcc11 | 350.577337 |
| mfcc19 | 1150.719005 | mfcc10 | 325.449136 |
| mfcc1 | 953.097621 | mfcc21 | 161.538421 |
| mfcc13 | 898.988289 | mfcc22 | 159.591179 |
| mfcc18 | 762.434767 | mfcc75 | 68.571325 |
| mfcc3 | 722.572833 | mfcc128 | 10.697854 |

According to Table 1, Spectrum Bandwidth has the maximum correlation coefficient (0.492264) values with the target variable (label) where Mel Frequency Cepstral Coefficient- mfcc128 has the lowest correlation coefficient value with the target variable. Besides, Roll Off, Spectral Centroid and Zero Crossing Rate have positive relation with the target variable. Another worth mentioning point, the Mel Frequency Cepstral Coefficient- mfcc20 has a positive relation (0.358174) with the target variable; however, mfcc21, mfcc22, mfcc75 and mfcc128 have the negative relation and these negative scores are gradually increase from mfcc21 to mfcc128.

Based on Table 2, Roll Off has the maximum Select_K_Best scores (2446.838434) with the target variable (label). On the contrary, Mel Frequency Cepstral Coefficient- mfcc128 has the lowest scores with the target variable. Spectral Bandwidth (spectral bandwidth) and Spectral

Centroid (spectral centroid) has significant scores. Mel Frequency Cepstral Coefficient's mfcc14, mfcc6, mfcc8, mfcc2, mfcc7, mfcc4 has good scores (more than Zero Crossing Rate, Root Mean Square Error and Chromogram). Also, mfcc18, mfcc19 and mfcc20 have citable scores (better than Root Mean Square Error and Chromogram). Nevertheless, the Select_K_Best scores are decreased notably after mfcc20. Table 1 and 3 concludes that the most preferred 26 features for this research are Spectral Bandwidth, Spectral Centroid, Roll Off, Zero Crossing Rate, Root Mean Square Error, Chromagram, and Mel Frequency Cepstral Coefficients' sequence- mfcc1 to mfcc20.

## 3.4. Implement Machine Learning Techniques

Different deep and non-deep machine learning techniques are applied to the extracted features and their performance are evaluated. Three deep learning techniques namely Sequential Neural Network, Convolutional Neural Network, and Recurrent Neural Network are implemented. Moreover, four different non-deep learning techniques including Support Vector Machine, Random Forest, Naïve Bayes and Decision Tree are implemented and their implementation details are shared from the conventional python-based library modules. Our contribution includes to design the most suitable architectures of the deep learning techniques and thus, we explain the implementation details of the deep learning techniques only. Before implementing all the machine learning techniques, the dataset split into 80:20 ratio for the training and testing datasets. A similar number of features (26 features) are preferred for implementing these classification techniques. Besides, 10-fold cross-validation are applied for all these techniques.

### 3.4.1. Sequential Neural Network (Sequential NN)

Sequential NN is a computational neural network technique that consists of multiple processing variables that receives inputs and provides outputs based on their predefined optimizer and activation function. Sequential NN is configured with the input layer, 4 fully connected dense layers with 256, 128, 64 and 1 neuron(s) and output layer. Figure 7 presents the model. Besides, the optimizer- AdaMax and activation function- Sigmoid is used to apply Sequential NN technique in this research.

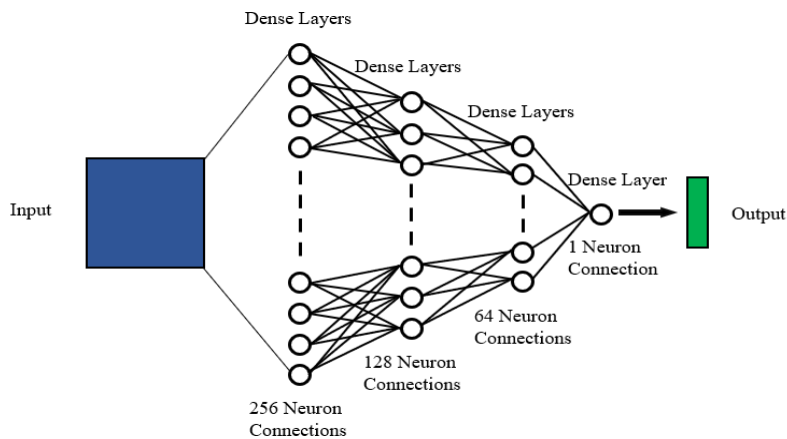

Figure 7.  The configuration of Sequential Neural Network

### 3.4.2. Convolutional Neural Network (CNN)

CNN is an effective and significant Artificial Neural Network technique which is used in the multilevel data processing. For observing and analyzing the performance of CNN, the optimizer-Sigmoid, activation function- AdaMax, 1000 epochs, and 128 batch size are used in our configuration. Figure 8 presents the architecture of the model. We use the same Sequential NN technique from Keras and add a 2D Conv Layer with 64 filters and (8,8) grid size of the kernel, and a Sigmoid activation function. We specify the 3D input shape into X train consists of 26 feature values with a mono dimension channel. After the convolution process, the feature maps are 20 x 64 using the following formulation: feature map size = (input features size – kernel size + stride size) x filter size. Then we add a max-pooling 2D layer with (2, 2) pool/grid size, (1,1) stride size (default value), and zero paddings. Pooled Feature Maps = feature map size/pooled size) x filter size = 20/2 x 64=10x64. Then the Pooled Feature Maps are flattened and propagated through four dense layers (256, 128, 64, 1).

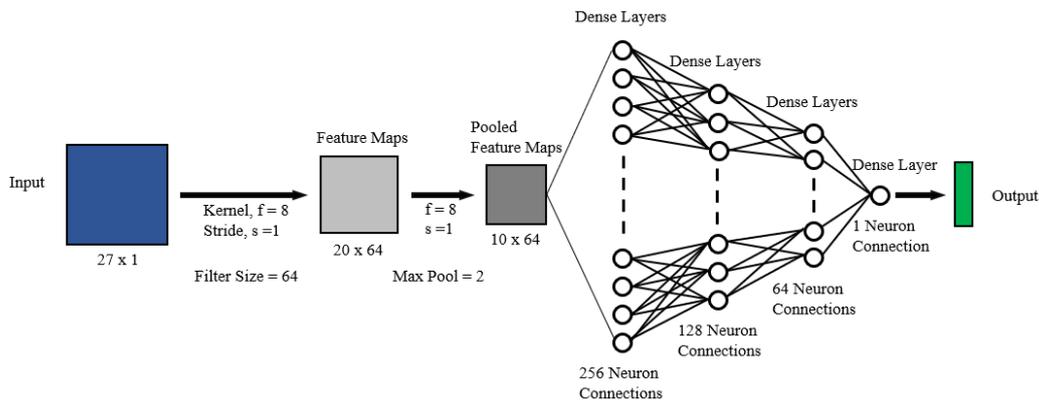

Figure 8. The configuration of Convolutional Neural Network

### 3.4.3. Recurrent Neural Network (RNN)

RNN has complicated structure and slow training pace. In addition, RNN is invariant to scale and rotation. RNN also has larger training time and consists vanishing gradient problem. RNN with LSTM are free from vanishing gradient problem thus we can explore more layers to improve the technique performance. We implement the same CNN architecture with additional two LSTM layers and presented in Figure 9. Besides, the fixed Learning Rate is 1e-3 and Decay is 1e-5 are implemented.

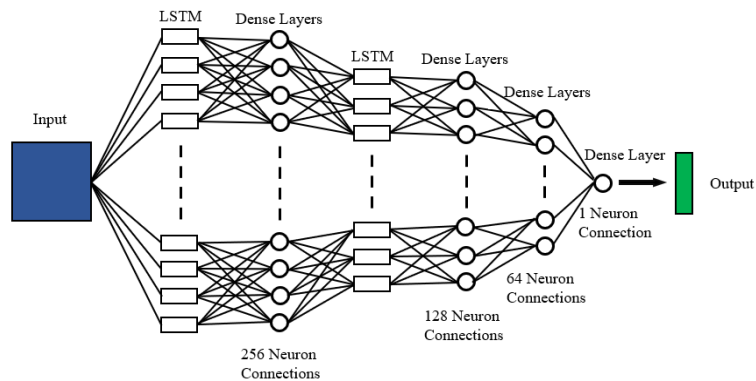

Figure 9. The configuration of Recurrent Neural Network

# 4. RESULT

This section states the performance analysis of the applied techniques and preferred features for this research. The performances are varied for the different techniques, features, input layers, activation functions, optimizers. Throughout the observation on outputs, the most compatible technique and convenient features are identified for this research.

## 4.1. Performance Analysis of Deep Learning Techniques

Table 3. Performance of Sequential NN based on different activation functions and optimizers.

| Optimizer<br><br>Activator | Adam | Ada Delta | Ada Grad | Ada Max | FTRL | Nadam | RMS prop | SGD |
|---|---|---|---|---|---|---|---|---|
| Relu | 98.40% | 93.73% | 98.16% | 98.03% | 71.74% | 98.52% | 99.01% | 98.03% |
| Sigmoid | 97.79% | 72.72% | 89.56% | 99.26% | 75.68% | 98.03% | 98.16% | 90.42% |
| Tanh | 97.79% | 93.37% | 96.68% | 98.28% | 85.14% | 97.79% | 98.53% | 98.16% |

At first, we need to find out the best architecture for Sequential NN. There are three activation functions - Relu (Rectified Linear Unit), Sigmoid, and Tanh (Hyperbolic Tangent Activation Function)-are available in Python's Keras library. Besides, eight optimizers - Adam (Adaptive Moment Estimation), AdaDelta, Ada- Grad, AdaMax, FTRL (Follow the Regularized Leader), Nadam (Nesterov-accelerated Adaptive Moment Estimation), RMSProp (Root Mean Squared Propagation), SGD (Stochastic Gradient Descent) – are available in Keras also. Therefore, this research needs to choose the best activation function and optimizer from the above- mentioned activation functions and optimizers list for performance analysis. According to Table 3, Sequential NN's highest performance- 99.26% accuracy is obtained with the combination of the Sigmoid activation function and AdaMax optimizer. The lowest performance achieves 71.74% accuracy by the FTRL activation function and Relu optimizer. Significant performance through the AdaMax optimizer using Relu (98.03% accuracy) and Tanh (98.28% accuracy) activator function. Since the optimizer- AdaMax and activation function- Sigmoid have showed the best performance and the combination is recommended for future use for all deep learning techniques. We also need to investigate the optimum number of features. Although the features- mfcc21 to mfcc128 have negative values by Correlation Coefficient and low scores by Select_K_Best algorithm but for better observation and analysis, we implement the Sequential NN with 26 features (without mfcc21 to mfcc128) and then with 27 features (by adding mfcc21) and thereafter 28 features (by adding mfcc22 later) with the Sigmoid activation function and AdaMax optimizer. The accuracy with 26 features is 99.28%, with 27 features the accuracy is 93.26% and with 28 features the accuracy is 90.71%. The overall performance is decreased with the increment of additional features than 26. Thus, 26 features are selected for further experiments.

After that, the deep learning techniques are applied with training and testing datasets with fixed epochs 1000. Table 4 presents the applied techniques' performance. The Sequential NN achieves the highest accuracy- 99.28% among the three deep learning techniques. Following the Sequential NN, the RNN technique performs- 93.57% accuracy. Besides, CNN obtains 85.04% accuracy and that is the lowest accuracy among the techniques. Figure 10(a), Figure 10(b) and Figure 10(c) shows the performance learning graphs of the Sequential NN, RNN and CNN, respectively.

Table 4. Performance of deep learning techniques.

| Techniques | Performance (accuracy) |
|---|---|
| Sequential Neural Network | 99.28% |
| Recurrent Neural Network | 93.57% |
| Convolutional Neural Network | 85.04% |

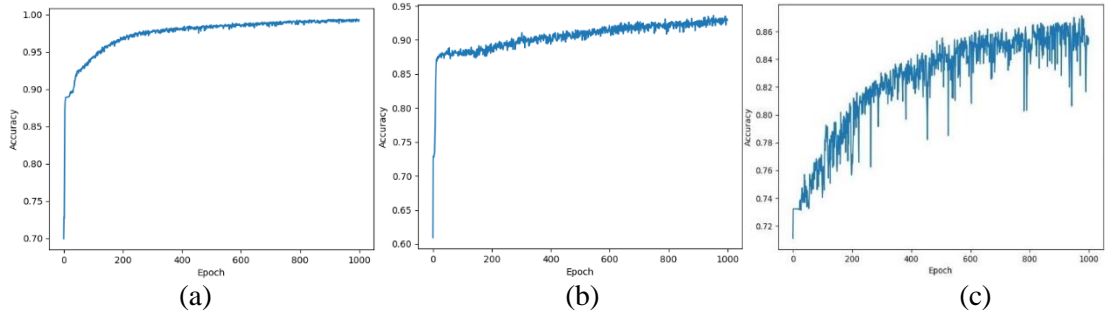

(a)                              (b)                              (c)

Figure 10. The Learning Graph of- (a) Sequential NN, (b) RNN, (c) CNN

Usually, Sequential NN plays notable performance with feature based numerical data sets, RNN plays better in text sequential datasets. However, CNN plays an important role to classify with image maps rather than numerical datasets. Table 4 reflects the similar results according to their usual characteristics. We extend our experiments to observe the image-based performance of Convolution Neural Network. We generate twenty (20) MFCC wave spectrum images from each wave file. The generated images' pixel size is 1440 x 1920 (width x height) with 300 dpi initially. Nevertheless, that is too time-consuming to apply the CNN based on these large pixel-sized images (4100 images). Therefore, we reduce the pixel size from 1440 x 1920 to 384 x 512 for all images and implement the CNN on those images. The similar configuration (the training set and testing set ratio 80:20, Sigmoid optimizer, AdaMax as activation function, 1000 epochs, batch size- 128, 4 layers where the primary layer formed with 256 neurons connection) is preferred in this implementation also. After this execution, CNN achieves 74.32% accuracy, which is less than CNN's previous performance – 85.04% accuracy (based on the dataset). More notably, it denotes that CNN performances (dataset-based and image-based) are less than other deep learning techniques. Thus, we are concluding that among the deep learning techniques Sequential NN performs better for feature-based bee data classification.

## 4.2. Performance Analysis of Non-Deep Learning Techniques

For proposing empirical research, four non-deep learning techniques (also known as classification techniques) – Support Vector Machine, Random Forest, Naïve Bayes and Decision Tree - are applied in this experiment. Typically, SVM measures the performance by mapping data to a high-dimensional feature space so that the data labels can be categorized where the NB follows the probabilistic method to provide the output. On the contrary, DT formulates the classification and generates the output by the Gini Index and Entropy, and RF determines performance through Gini Importance, where a couple of combined formulas are integrated. Following the deep learning techniques, a similar number of features (26 features) are preferred for these classification techniques. Besides, 10-fold cross-validation is applied with all techniques. Table 5 shows the performance of these implementations.

Table 5. Performance of non-deep learning techniques.

| Techniques | Performance (accuracy) |
|---|---|
| Support Vector Machine | 92.65% |
| Random Forest | 97.74% |
| Naïve Bayes | 88.06% |
| Decision Tree | 94.72% |

Table 5 presents that Random Forest achieves the highest accuracy- 97.74% as a non-deep learning technique; a more effective non-deep learning technique to classify bee and non-bee sounds. Following the Random Forest, the Decision Tree and Support Vector Machine has citable results. On the other hand, Naïve Bayes has 88.06% accuracy, which is the lowest performance based on applied non-deep learning techniques. The following section describes the performance comparison between deep learning and non-deep learning techniques to classify beehive sounds.

## 4.3. Performance Comparison Between Deep Learning Techniques and Non-Deep Learning Techniques

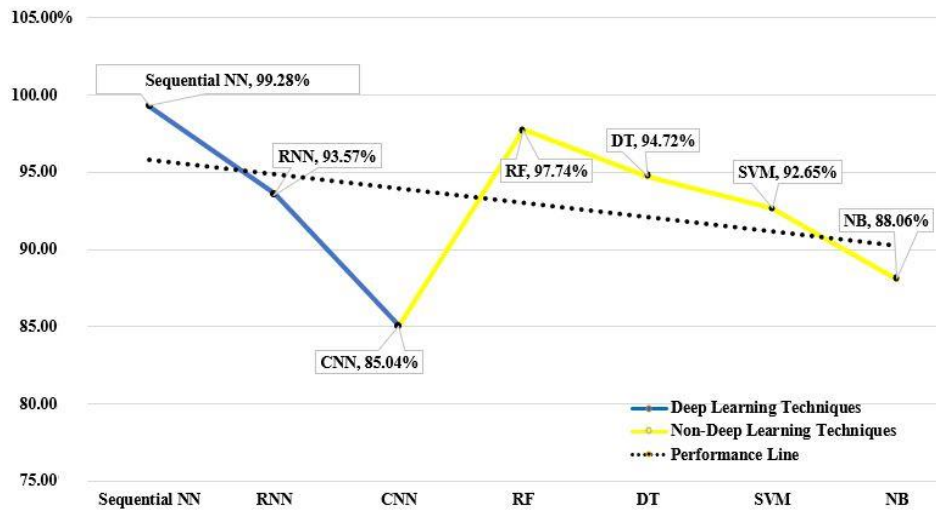

Figure 11. Performance comparison between deep learning techniques and non-deep learning techniques

Figure 11 demonstrates the comparison of performance between deep learning technique and non-deep learning techniques by the graph with liner lines. According to this figure, the Sequential NN has the highest performance (99.28% accuracy). Following the Sequential NN, the Random Forest achieves the second-highest performance (97.74% accuracy). Besides, The Random Forest and Decision Tree (94.72% accuracy) have better results than the Recurrent Neural Network (93.57% accuracy). Similarly, the Support Vector Machine (92.65% accuracy) and Naïve Bayes (88.06% accuracy) have better results than the Convolutional Neural Network (85.04% accuracy).

Besides, the linear trendline or performance line shows the overall direction or general pattern of the implemented techniques. Both axes' (non-deep learning and deep learning techniques) directions are downward according to their techniques' performance. However, through the trendline analysis, we observe that the performance line is declining from deep learning techniques to the non-deep learning technique. This observation indicates that deep learning techniques are more relevant in beehive sound identification.

## 4.4. Research Model's Performance Validation

For justifying this research significance, we conduct validity testing on exceptional data. Through this validation process, we observe the adaptability of the unknown sounds in this experimental model. Thus, we extracted randomly five sound clips - bee, non-bee and combined (bee and non-bee) - sound from the source file's different portions and predicted the output by preferred techniques; however, the training data remained the same. The five wave files, namely wavefile1: full 2 seconds bee's sound, wavefile2: full 2 seconds nobee's sound, wavefile3: 1-second bee's sound with 1-second nobee's sound, wavefile4: 1.25 seconds bee's sound with 0.75-second nobee's sound, and wavefile5: 1.25 seconds nobee's sound with 0.75-second bee's sound. From those five wave files, the data from 26 preferred features are computed and predicted their labels (bee's sound or nobee's sound) based on the applied techniques. The wave- file3 is a combined form of 1-second bee's with 1-second nobee's sound; thus, we consider this one as a don't care state, and any sound type of bee or nobee is considered as the correct prediction. For wavefile4 and wavefile5, we consider the correct labels by the maximum sound (bee/nobee) duration in those wave files; therefore, the wavefile4 and wavefile5 are considered bee and nobee, respectively, as their correct prediction. Table 6 shows the validation of prediction matching accuracy based on implemented techniques.

Table 6. Validation of the applied techniques with their prediction matching accuracy.

| Wave File Names | Non-Deep Learning Techniques Prediction | | | | Deep Learning Techniques Prediction | | | Original Wave File (Label) |
|---|---|---|---|---|---|---|---|---|
| | SVM | RF | NB | DT | Sequential NN | RNN | CNN | |
| **wavefile1** | bee | bee | bee | bee | bee | bee | bee | 2.00 sec bee |
| **wavefile2** | nobee | nobee | nobee | nobee | nobee | nobee | bee | 2.00 sec nobee |
| **wavefile3** | bee | bee | bee | nobee | nobee | nobee | nobee | 1.00 sec bee + 1.00 sec nobee |
| **wavefile4** | bee | nobee | bee | nobee | bee | nobee | bee | 1.25 sec bee + 0.75 sec nobee |
| **wavefile5** | bee | bee | bee | nobee | nobee | nobee | nobee | 0.75 sec bee + 1.25 sec nobee |
| **Techniques' Prediction Matching Accuracy** | 80% | 60% | 80% | 80% | 100% | 80% | 80% | |

This validation process illustrates the prediction performance of non-deep learning and deep learning techniques, where deep learning techniques show notable outcomes. The deep learning techniques have achieved maximum accuracy of 80% (by Support Vector Machine, Naïve Bayes and Decision Trees) to predict the bee or non-bee sounds, whereas the lowest performance- 80% for the deep learning techniques (by Recurrent Neural Network and Convolutional Neural Network) of prediction those sounds' categories. However, the Sequential Neural Network performs significantly better than other techniques (has achieved 100% accuracy); it predicts all waves files to detect the bee and non-sound exactly.

## 5. LIMITATIONS AND FUTURE WORK

The limitation in our system is that we have to execute the annotation system manually through the Sonic Visualiser3 tool. As we are implementing supervised classifiers; therefore, the performance of the classifiers is entirely dependent on the quality of the annotations. A single incorrect annotation will cause the whole system to malfunction in the real scenario. For this reason, in the near future, we are planning to integrate an automated annotation system based on our Sequential Neural Network. We will prefer to fix a specific window size (between 1 and 2 seconds) to generate waves, which will then be sent to an ANN for classification. The selection on an annotation will be made based on the results of the ANN analysis. The optimum window size will be another direction for our future research.

## 6. CONCLUSION

In this work, we create our own datasets from the annotated datasets of past research. In total, 134 features are selected in the initial phase. Correlation Coefficient and Select_K_Best are two feature engineering approaches applied to select the most 26 suitable features. We explore both deep learning and non-deep learning machine learning approaches on the 26 features datasets and observe their accuracy. Sequential NN with Sigmoid activation function and AdaMax optimizer performs best among the other deep learning techniques and achieves 99.28% accuracy. Random forest performs best among the non-deep learning techniques with 97.74% accuracy. Sequential NN is also performing best in the combined data as well. The model performs accurately with up to 25% combined sounds and is recommended to use in the future. The research also highlights that RNN shows a better result to classify bee sounds from the non-beehive noises compared to CNN but not good as Sequential NN. This is a common scenario as Sequential NN performs well in tabular type datasets, RNN also shows a satisfactory performance in time series, text, or audio datasets, and CNN performs well in image datasets. Our next research will explore the audio signal directly to RNN, and CNN will check their performance.

## AUTHORS

**Shah Jafor Sadeek Quaderi** pursued M.S in Applied Computing from the University of Malaya, Malaysia, in 2021 and B.S in Computer Science and Engineering from East West University Bangladesh in 2018. He was Graduate Research Assistant in Artificial Intelligence Lab, University of Malaya, Kuala Lumpur, Malaysia. He served as a Research Assistant in AISIP Lab, Bangladesh, from 2018 to 2019 and worked as an Undergraduate Teaching Assistant during his B.S degree. His preferred research areas- Information Retrieval, Natural Language Processing, Data Analysis, Prediction Model, Machine Learning and Digital Circuits.

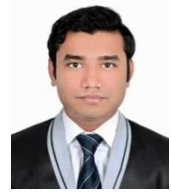

**Sadia Afrin Labonno** is an undergraduate student and researcher at the AISIP lab. Her research interest includes AI, Machine Learning Models, and Data Mining Algorithms.

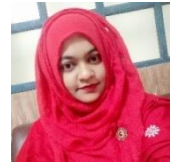

**Sadia Mostafa** is an undergraduate student and researcher at the AISIP lab. Her research interest includes Machine Learning Models, and Big-Data Applications.

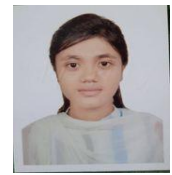

**Shamim Akhter** received the B.S. in computer science from the American International University of Bangladesh, in 2001, the M.S in computer science and information management from the Asian Institute of Technology, Thailand, in 2005 and the Ph.D. degree in information processing from Tokyo Institute of Technology, Japan, in 2009. From 2009 to 2011, he was a JSPS post-doctoral research fellow with the National Institute of Informatics, Japan. Currently he is affiliated with Stamford University Bangladesh as Professor. He is the author of a book and more than 60 articles. His research interests include Machine Intelligence, Intelligent Algorithms, Evolutionary Algorithms and Parallelization Models.

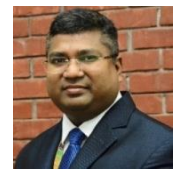